\documentclass{PoS}
\usepackage[svgnames]{xcolor} % Specify colors by their 'svgnames', for a full list of all colors available see here: http://www.latextemplates.com/svgnames-colors

\usepackage{times} % Use the times font
\usepackage{amsfonts, amsmath, amsthm, amssymb} % For math fonts, symbols and environments
\title{Neutrino spin and spin-flavor oscillations \\ in matter currents and magnetic fields}

\ShortTitle{Neutrino spin oscillations in magnetic fields and matter currents}

\author{Pavel Pustoshny\\
        Department of Theoretical Physics, Lomonosov Moscow State University, 119992 Moscow, Russia\\
        E-mail: \email{pg.pustoshnyy@physics.msu.ru}}

\author{Vadim Shakhov\\
        Department of Theoretical Physics, Lomonosov Moscow State University, 119992 Moscow, Russia\\
        E-mail: \email{schakhov.vv15@physics.msu.ru}}

\author{\speaker {Alexander Studenikin}\\
        Department of Theoretical Physics, Lomonosov Moscow State University, 119992 Moscow, Russia\\
        Joint Institute for Nuclear Research, 141980 Dubna, Moscow Region, Russia\\
        E-mail: \email{studenik@srd.sinp.msu.ru}}

\abstract{After a brief  history of two known types of neutrino mixing and oscillations, including neutrino spin and spin-flavour oscillations in the transversal magnetic field, we perform systematic study of  a new phenomenon of neutrino spin and spin-flavour oscillations engendered by the transversal matter currents on the bases of the developed quantum treatment of the phenomenon. Possibilities for the resonance amplification of these new types of oscillations by the longitudinal matter currents and longitudinal magnetic fields are analyzed. We also consider modifications of the oscillation probabilities due to possible arbitrary orientation of the magnetic field vector ${\bf B}$ and the matter velocity ${\bf v}$.}

\FullConference{
European Physical Society Conference on High Energy Physics - EPS-HEP2019 -\\
			10-17 July, 2019\\
			Ghent, Belgium}

\begin{document}

\section* {The history of neutrino spin oscillations in transversal
matter currents and/or transversally polarized matter}

For many years, until  2004, it was believed that a neutrino helicity precession
and the corresponding spin oscillations can be induced by the neutrino magnetic
interactions with an external electromagnetic field that provided the existence of the transversal magnetic field component ${\bf B}_{\perp}$ in the particles rest frame.
A new and very interesting possibility for neutrino spin
(and spin-flavour) oscillations engendered by the neutrino interaction with matter background was
proposed and investigated for first time in \cite{Studenikin:2004bu}. It
was shown \cite{Studenikin:2004bu} that neutrino spin oscillations
can be induced not only by the neutrino interaction with a  magnetic field, as it was believed
before, but also by neutrino interactions with matter in the case when there
is a transversal matter current or matter polarization. This new effect has been
explicitly highlighted in \cite{Studenikin:2004bu} :

{\it ``The possible emergence of neutrino spin oscillations  owing to neutrino
interaction with matter under the condition that there exists a nonzero transverse current component or
matter polarization is the most important new effect that follows from the investigation
of neutrino-spin oscillations in Section 4. So far, it has been assumed that neutrino-spin oscillations
may arise only in the case where there exists a nonzero transverse magnetic field in the neutrino rest
frame." }

For historical notes reviewing studies of the discussed effect see in \cite{Studenikin:2016iwq, Studenikin:2016ykv, Studenikin:NOW_2016}.
It should be noted that the predicted effect exist regardless of a source of the background matter transversal current or polarization (that can be a background magnetic field, for instance). It is also interesting to note th
the significant change in the neutrino (spin and flavour) oscillations
pattern due to only the longitudinal relativistic motion of matter was indicated as one of four new effects discussed in  \cite{Studenikin:2004tv}.

Note that the existence of the discussed effect of neutrino spin oscillations engendered by the transversal matter current and matter polarization and its possible impact in astrophysics have been confirmed in a series of papers \cite{Cirigliano:2014aoa, Volpe:2015rla, Kartavtsev:2015eva, Dobrynina:2016rwy}. In our recent paper \cite{Pustoshny:2018jxb} we have developed a consistent quantum treatment of the neutrino spin and spin-flavor oscillations engendered by the transversal matter currents. The presence of the transversal and longitudinal magnetic fields as well as the longitudinal matter currents are accounted for. The developed treatment \cite{Pustoshny:2018jxb} also allows to account neutrino nonstandard interactions. In addition, different possibilities for the resonance amplification of these kind of neutrino spin and spin-flavor oscillations are considered. Here below \cite{Shakhov:2019} we consider a possibility of different orientation of the transversal matter current $j_{\perp}$ in respect to the vector of the transversal magnetic field $B_{\perp}$.

\color{Black}
%SEMICLASSICAL TREATMENT
\section* {Neutrino spin oscillations  $ \nu_{e}^L~\Leftarrow~(j_{\perp})~\Rightarrow~\nu_{e}^R$ engendered by transversal matter currents: semiclassical treatment}
Following the discussion in \cite{Studenikin:2004bu} consider, as an example,  an electron neutrino spin precession in the case when neutrinos with the Standard Model interaction are propagating through moving and polarized matter composed of electrons (electron gas) in the presence of an electromagnetic field given by the electromagnetic-field tensor $F_{\mu \nu}=({\bf E}, {\bf B})$.
To derive the neutrino spin oscillation probability in the transversal matter current we use the generalized Bargmann-Michel-Telegdi equation that describes  the evolution of the
three-di\-men\-sio\-nal neutrino spin vector $\bf S $,
\begin{equation} \label{S}
\frac{d {\bf S}}{dt} =
\frac{2}{\gamma} \left[ {\bf S} \times \left({\bf B}_0+{\bf M}_0\right) \right],
\end{equation}
where the magnetic field ${\bf B}_0$ in the neutrino rest frame is determined by the transversal and longitudinal (with respect to the neutrino motion) magnetic and electric field components in the laboratory frame,
\begin{equation}
{\bf B}_{0}=\gamma\left(\mathbf{B}_{\perp}+\frac{1}{\gamma} \mathbf{B}_{ \|}+\sqrt{1-\gamma^{-2}}\left[\mathbf{E}_{\perp} \times \frac{\boldsymbol{\beta}}{\beta}\right]\right)
\end{equation}
$\gamma=\left(1-\beta^{2}\right)^{-\frac{1}{2}}$, $\boldsymbol{\beta}$ is the neutrino velocity. The matter term $\mathbf{M}_0$ in Eq. (\ref{S}) is also  composed of the transversal $\mathbf{M}_{0_{\perp}}$ and longitudinal $\mathbf{M}_{0_{||}}$ parts.
\begin{equation}
\mathbf{M}_{0}=\mathbf{M}_{0_{||}}+\mathbf{M}_{0_{\perp}}
\end{equation}
\begin{equation}
\begin{array}{c}{\mathbf{M}_{0_{ \|}}=\gamma \boldsymbol{\beta} \frac{n_{0}}{\sqrt{1-v_{e}^{2}}}\left\{\rho_{e}^{(1)}\left(1-\frac{\mathbf{v}_{e} \boldsymbol{\beta}}{1-\gamma^{-2}}\right)\right\}-} \\ {-\frac{\rho_{e}^{(2)}}{1-\gamma^{-2}}\left\{\boldsymbol{\zeta}_{e} \boldsymbol{\beta} \sqrt{1-v_{e}^{2}}+\left(\boldsymbol{\zeta}_{e} \mathbf{v}_{e} \frac{\left(\boldsymbol{\beta} \mathbf{v}_{e}\right)}{1+\sqrt{1-v_{e}^{2}}}\right)\right\},}\end{array}
\end{equation}
\begin{equation} \label{M_0_perp}
\begin{array}{c}
{\mathbf{M}_{0_{\perp}}=-\frac{n_{0}}{\sqrt{1-v_{e}^{2}}}
\mathbf{v}_{e_{\perp}}\left(\rho_{e}^{(1)}+\rho_{e}^{(2)} \frac{\left(\zeta_{e} \mathbf{v}_{e}\right)}{1+\sqrt{1-v_{e}^{2}}}\right)+} \\
{+\boldsymbol{\zeta}_{e_{\perp}} \rho_{e}^{(2)} \sqrt{1-v_{e}^{2}}.}
\end{array}
\end{equation}
Here $n_{0}=n_{e} \sqrt{1-v_{e}^{2}}$ is the invariant number density of matter given in the reference frame for which the total speed of matter is zero. The vectors $\mathbf{V}_{e},$ and $\boldsymbol{\zeta}_{e}\left(0 \leq\left|\zeta_{e}\right|^{2} \leq 1\right)$ denote, respectively, the speed of the reference frame in which the mean momentum of matter (electrons) is zero, and the mean value of the polarization vector of the background electrons in the above mentioned reference frame. The coefficients $\rho_{e}^{(1,2)}$ calculated withinthe extended Standard Model supplied with SU(2)-singlet right-handed neutrino $\nu _R$ are respectively, $\rho_{e}^{(1)}=\frac{\tilde{G}_{F}}{2 \sqrt{2} \mu}, \quad \rho_{e}^{(2)}=-\frac{G_{F}}{2 \sqrt{2} \mu},$ where $\tilde{G}_{F}~=~G_{F}\left(1+4 \sin ^{2} \theta_{W}\right).$

For neutrino evolution between two neutrino states $\nu_{e}^{L} \Leftrightarrow \nu_{e}^{R}$ in presence of the magnetic field and moving matter we get \cite{Studenikin:2004bu} the following equation
\begin{equation}
i \frac{d}{d t} \begin{pmatrix} \nu _e^L \\ \nu _e^R \end{pmatrix} = \mu
\begin{pmatrix}
{\frac{1}{\gamma}|\mathbf{M}_{0 \|}+\mathbf{B}_{0 \|}|} & {|\mathbf{B}_{\perp}+\frac{1}{\gamma} \mathbf{M}_{0 \perp}|} \\
{|\mathbf{B}_{\perp}+\frac{1}{\gamma} \mathbf{M}_{0 \perp}|} & {-\frac{1}{\gamma}|\mathbf{M}_{0 \|}+\mathbf{B}_{0 \|}|}
\end{pmatrix}
\begin{pmatrix} \nu _e^L \\ \nu _e^R \end{pmatrix}.
\end{equation}
Thus, the probability of the neutrino spin oscillations in the
adiabatic approximation is given by
\begin{equation} \label{prob}
\begin{array}{c} {P_{\nu_{e}^{L} \rightarrow \nu_{e}^{R}}(x) =\sin^22\theta_{\mathrm{eff}}\sin ^{2} \frac{\pi x}{L_{\mathrm{eff}}},\ \ \
\sin^22\theta_{\mathrm{eff}} = \frac{E_{\mathrm{eff}}^{2}}{E_{\mathrm{eff}}^{2}+\Delta_{\mathrm{eff}}^{2}}},\\
{L_{\mathrm{eff}} =\frac{\pi}{\sqrt{E_{\mathrm{eff}}^{2}+\Delta_{\mathrm{eff}}^{2}}}}, \end{array}
\end{equation}
where
\begin{equation}
E_{\mathrm{eff}} = \mu |\mathbf{B}_{\perp}+\frac{1}{\gamma} \mathbf{M}_{0 \perp}| \ \ \ \Delta_{\mathrm{eff}} = \frac{\mu}{\gamma} |\mathbf{M}_{0 \|}+\mathbf{B}_{0 \|}|.
\end{equation}
From this it follows \cite{Studenikin:2004bu} that even in the absence of the transversal magnetic field the neutrino spin oscillations can appear due to the transversal matter current.
\color{Black}
\section*{Neutrino spin oscillations  $\nu_{e}^L~\Leftarrow~(j_{\perp})~\Rightarrow~\nu_{e}^R$ engendered by transversal matter currents: quantum treatment}
Here below we continue our studies of the effect of neutrino spin evolution induced by the transversal matter currents and develop a consistent derivation of the effect based on the direct calculation of the spin evolution effective Hamiltonian in the case when a neutrino is propagating in matter transversal currents.

Consider two flavour neutrinos with two possible helicities $\nu_{f}=\left(\nu_{e}^{+}, \nu_{e}^{-}, \nu_{\mu}^{+}, \nu_{\mu}^{-}\right)^{T}$  in moving matter composed of neutrons. The neutrino interaction Lagrangian reads
\begin{equation}
L_{i n t}=-f^{\mu} \sum_{l} \overline{\nu}_{l}(x) \gamma_{\mu} \frac{1+\gamma_{5}}{2} \nu_{l}(x) = -f^{\mu} \sum_{i} \overline{\nu}_{i}(x) \gamma_{\mu} \frac{1+\gamma_{5}}{2} \nu_{i}(x),
\end{equation}
where $f^{\mu}=-\frac{G_{F}}{\sqrt{2}} n(1, \mathbf{v}),$ $l=e,\mu$ indicates the neutrino flavour, $i=1,2$ indicates the neutrino mass state and the matter potential $f^{\mu}$ depends on the velocity of matter $\mathbf{v}=(v_1,v_2,v_3)$ and on the neutron number density in the laboratory reference frame $n=\frac{n_0}{\sqrt{1-v^2}}$. Each of the flavour neutrinos is a superposition of the neutrino mass states,
\begin{equation}
\nu_{e}^{ \pm}=\nu_{1}^{ \pm} \cos \theta+\nu_{2}^{ \pm} \sin \theta, \quad \nu_{\mu}^{ \pm}=-\nu_{1}^{ \pm} \sin \theta+\nu_{2}^{ \pm} \cos \theta.
\end{equation}
The neutrino evolution equation in the flavour basis is
\begin{equation}
i \frac{d}{d t} \nu_{f}=\left(H_{0}^{e f f}+\Delta H^{e f f}\right) \nu_{f},
\end{equation}
where the effective Hamiltonian consists of the vacuum and interaction parts:
\begin{equation}
H^{e f f}=H_{0}^{e f f}+\Delta H^{e f f}.
\end{equation}
$\Delta H^{e f f}$ can be expressed as \cite{Studenikin:2016ykv}
\begin{equation}
\Delta H^{e f f}=\left(\begin{array}{cccc}{\Delta_{e e}^{++}} & {\Delta_{e e}^{+-}} & {\Delta_{e \mu}^{++}} & {\Delta_{e \mu}^{+-}} \\ {\Delta_{e e}^{-+}} & {\Delta_{e e}^{-}} & {\Delta_{e \mu}^{-+}} & {\Delta_{e \mu}^{--}} \\ {\Delta_{\mu e}^{++}} & {\Delta_{\mu e}^{+-}} & {\Delta_{\mu \mu}^{++}} & {\Delta_{\mu \mu}^{+-}} \\ {\Delta_{\mu e}^{-+}} & {\Delta_{\mu e}^{-}} & {\Delta_{\mu \mu}^{-+}} & {\Delta_{\mu \mu}^{--}}\end{array}\right)
\end{equation}
where
\begin{equation}
\Delta_{k l}^{s s^{\prime}}=\left\langle\nu_{k}^{s}\left|H_{i n t}\right| \nu_{l}^{s^{\prime}}\right\rangle
\end{equation}
$k, l=e, \mu,$ $s, s^{\prime}=\pm.$ In evaluation of $\Delta_{k l}^{s s^{\prime}}$ we have first introduced the neutrino flavour states $\nu_{k}^{s}$ and $\nu_{l}^{s^{\prime}}$ as superpositions of the mass states $\nu_{1,2}^{\pm}$. Then, using the exact free neutrino mass states spinors,
\begin{equation}
\nu_{\alpha}^{s}=C_{\alpha}\left(\begin{array}{c}{u_{\alpha}^{s}} \\ {\frac{\sigma p_{\alpha}}{E_{\alpha}+m_{\alpha}} u_{\alpha}^{s}}\end{array}\right) \sqrt{\frac{\left(E_{\alpha}+m_{\alpha}\right)}{2 E_{\alpha}}} \exp \left(i \boldsymbol{p}_{\boldsymbol{\alpha}} \boldsymbol{x}\right)
\end{equation}
where the two-component spinors define neutrino helicity states, and are given by
\begin{equation}
u_{\alpha}^+=\left(\begin{array}{l}{1} \\ {0}\end{array}\right), \quad u_{\alpha}^-=\left(\begin{array}{l}{0} \\ {1}\end{array}\right)
\end{equation}
for the typical term $\Delta_{\alpha \alpha^{\prime}}^{s s^{\prime}}=\left\langle\nu_{\alpha}^{s}\left|\Delta H^{S M}\right| \nu_{\alpha^{\prime}}^{s^{\prime}}\right\rangle,$ that by fixing proper values of $\alpha, s, \alpha^{\prime}$ and $s^{\prime}$ an reproduces all of the elements of the neutrino evolution Hamiltonian $\Delta H^{e f f}$ that accounts for the effect of matter motion, we obtain \cite{Studenikin:2016ykv}
\begin{equation}
\Delta_{\alpha \alpha^{\prime}}^{s s^{\prime}}=\tilde{G} n\left\{u_{\alpha}^{s T}\left[\left(\begin{array}{cc}{0} & {0} \\ {0} & {2}\end{array}\right) v_{ \|}+\left(\begin{array}{cc}{0} & {\gamma_{\alpha}^{-1}} \\ {\gamma_{\alpha^{\prime}}^{-1}} & {0}\end{array}\right) v_{\perp}\right] u_{\alpha^{\prime}}^{s^{\prime}}\right\}\delta^{\alpha^{\prime}}_{\alpha}
\end{equation}
where $v_{||}$ and $v_{\perp}$ are the longitudinal and transversal velocities of the matter current, $\tilde{G}=\frac{G_{F}}{2 \sqrt{2}},\  n=\frac{n_{0}}{\sqrt{1-v^{2}}}$ and
\begin{equation}
\gamma_{\alpha \alpha^{\prime}}^{-1}=\frac{1}{2}\left(\gamma_{\alpha}^{-1}+\gamma_{\alpha^{\prime}}^{-1}\right), \quad \widetilde{\gamma}_{\alpha \alpha^{\prime}}^{-1}=\frac{1}{2}\left(\gamma_{\alpha}^{-1}-\gamma_{\alpha^{\prime}}^{-1}\right), \quad \gamma_{\alpha}^{-1}=\frac{m_{\alpha}}{E_{\alpha}}.
\end{equation}
so that the effective interaction Hamiltonian in the flavour basis has the following structure,
\begin{equation}
H^{eff}=n \tilde{G} \left( \begin{array}{cccc}
  {0}
& {\left(\frac{\eta}{\gamma}\right)_{e e} v_{\perp}}
& {0}
& {\left(\frac{\eta}{\gamma}\right)_{e \mu} v_{\perp}} \\
  {\left(\frac{\eta}{\gamma}\right)_{e e} v_{\perp}}
& {2\left(1-v_{\| }\right)}
& {\left(\frac{\eta}{\gamma}\right)_{e \mu} v_{\perp}}
& {0} \\
  {0}
& {\left(\frac{\eta}{\gamma}\right)_{e \mu} v_{\perp}}
& {0}
& {\left(\frac{\eta}{\gamma}\right)_{\mu \mu} v_{\perp}} \\
  {\left(\frac{\eta}{\gamma}\right)_{e \mu} v_{\perp}}
& {0}
& {\left(\frac{\eta}{\gamma}\right)_{\mu \mu} v_{\perp}}
& {2\left(1-v_{\| }\right)}\end{array}\right).
\end{equation}
Here we introduce the following formal notations:
\begin{equation}
\left(\frac{\eta}{\gamma}\right)_{e e} =\frac{\cos ^{2} \theta}{\gamma_{11}}+\frac{\sin ^{2} \theta}{\gamma_{22}},\quad
\left(\frac{\eta}{\gamma}\right)_{\mu \mu} =\frac{\sin ^{2} \theta}{\gamma_{11}}+\frac{\cos ^{2} \theta}{\gamma_{22}},\quad
\left(\frac{\eta}{\gamma}\right)_{e \mu} =\frac{\sin 2 \theta}{\tilde{\gamma}_{21}}.
\end{equation}

The flavor neutrino evolution Hamiltonian in the magnetic field $H^f_B$ can be calculated in the same way. One just should start from the neutrino electromagnetic interaction Lagrangian $L_{E M}~=~\frac{1}{2} \mu_{\alpha \alpha^{\prime}} \overline{\nu}_{\alpha^{\prime}} \sigma_{\mu \nu} \nu_{\alpha} F^{\mu \nu}~+~h . c .$
\begin{equation} \label{Ham}
H_{B}^{f} = \left( \begin{array}{cccc}{-\left(\frac{\mu}{\gamma}\right)_{e e} B_{\| }}
& {-\mu_{e e} B_{\perp}e^{-i\phi}}
& {-\left(\frac{\mu}{\gamma}\right)_{e \mu} B_{\| }}
& {-\mu_{e \mu} B_{\perp}e^{-i\phi}} \\
  {-\mu_{e e} B_{\perp}e^{i\phi}}
& {\left(\frac{\mu}{\gamma}\right)_{e e} B_{\| }}
& {-\mu_{e \mu} B_{\perp}e^{i\phi}}
& {\left(\frac{\mu}{\gamma}\right)_{e \mu} B_{\| }} \\
  {-\left(\frac{\mu}{\gamma}\right)_{e \mu} B_{\| }}
& {-\mu_{e \mu} B_{\perp}e^{-i\phi}}
& {-\left(\frac{\mu}{\gamma}\right)_{\mu \mu} B_{\| }}
& {-\mu_{\mu \mu} B_{\perp}e^{-i\phi}} \\
  {-\mu_{e \mu} B_{\perp}e^{i\phi}}
& {\left(\frac{\mu}{\gamma}\right)_{e \mu} B_{\| }}
& {-\mu_{\mu \mu} B_{\perp}e^{i\phi}}
& {\left(\frac{\mu}{\gamma}\right)_{\mu \mu} B_{\| }}
\end{array}\right),
\end{equation}
where $\phi$ is the angle between transversal components of magnetic field $\boldsymbol{B}_{\perp}$ and matter velocity $\mathbf{v}_{\perp}$ and
%ОБОЗНАЧЕНИЯ
\begin{equation}
\begin{aligned}\left(\frac{\mu}{\gamma}\right)_{e e} &=\frac{\mu_{11}}{\gamma_{11}} \cos ^{2} \theta+\frac{\mu_{22}}{\gamma_{22}} \sin ^{2} \theta+\frac{\mu_{12}}{\gamma_{12}} \sin 2 \theta, \\\left(\frac{\mu}{\gamma}\right)_{e \mu} &=\frac{\mu_{12}}{\gamma_{12}} \cos 2 \theta+\frac{1}{2}\left(\frac{\mu_{22}}{\gamma_{22}}-\frac{\mu_{11}}{\gamma_{11}}\right) \sin 2 \theta, \\\left(\frac{\mu}{\gamma}\right)_{\mu \mu} &=\frac{\mu_{11}}{\gamma_{11}} \sin ^{2} \theta+\frac{\mu_{22}}{\gamma_{22}} \cos ^{2} \theta-\frac{\mu_{12}}{\gamma_{12}} \sin 2 \theta, \end{aligned}
\end{equation}

%ОБОЗНАЧЕНИЯ
\begin{equation}
\begin{aligned} \mu_{e e} &=\mu_{11} \cos ^{2} \theta+\mu_{22} \sin ^{2} \theta+\mu_{12} \sin 2 \theta, \\ \mu_{e \mu} &=\mu_{12} \cos 2 \theta+\frac{1}{2}\left(\mu_{22}-\mu_{11}\right) \sin 2 \theta, \\ \mu_{\mu \mu} &=\mu_{11} \sin ^{2} \theta+\mu_{22} \cos ^{2} \theta-\mu_{12} \sin 2 \theta. \end{aligned}
\end{equation}
The Hamiltonian (\ref{Ham}) explicitly accounts \cite{Shakhov:2019} for a possibility of different orientation of the
transversal matter current $j_{\perp}$ in respect to the vector of the transversal magnetic field $B_{\perp}$.
%----------------------------------------------------------------------------------------
%   PROBABILITIES
%----------------------------------------------------------------------------------------
\color{Black}
\section*{Probability of neutrino spin-flavor oscillations $\nu_{e}^{L} \Leftarrow\left(j_{\perp}, B_{\perp}\right) \Rightarrow \nu_{\mu}^{R}$}
Consider two states of neutrino $(\nu^L_e,\nu^R_{\mu})$. The corresponding oscillations are governed by the evolution equation
\begin{equation}
\footnotesize
i \frac{d}{d t} \left( \begin{array}{c}
{\nu_{e}^{L}} \\ {\nu_{\mu}^{R}}\end{array}\right) =
\left( \begin{array}{cc}
{-\Delta M+\left(\frac{\mu}{\gamma}\right)_{e e} B_{\| }+\tilde{G} n(1-v \beta)} & -{\mu_{e \mu} B_{\perp}e^{i \phi}+\left(\frac{\eta}{\gamma}\right)_{e \mu} \tilde{G} n v_{\perp}} \\ -{\mu_{e \mu} B_{\perp}e^{-i \phi}+\left(\frac{\eta}{\gamma}\right)_{e \mu} \tilde{G} n v_{\perp}} & {\Delta M-\left(\frac{\mu}{\gamma}\right)_{\mu \mu} B_{ \|}-\tilde{G} n(1-v \beta)}\end{array}\right)
\left( \begin{array}{c}{\nu_{e}^{L}} \\
{\nu_{\mu}^{R}}\end{array}\right)
\end{equation}
For the oscillation $\nu_{e}^{L} \Leftarrow\left(j_{\perp}, B_{\perp}\right) \Rightarrow \nu_{\mu}^{R}$ probability we get (\ref{prob}) with
\begin{equation}
\begin{aligned}
E_{\mathrm{eff}} &=\sqrt{\left(-\mu_{e \mu} B_{\perp} \cos \phi+\left(\frac{\eta}{\gamma}\right)_{e \mu} \tilde{G} n v_{\perp}\right)^2 + \left(\mu _{e \mu} B_{\perp} \sin \phi\right)^2} \\
\Delta_{\mathrm{eff}} &=\left|\Delta M - \frac{1}{2}\left(\frac{\mu_{11}}{\gamma_{11}}+\frac{\mu_{22}}{\gamma_{22}}\right) B_{ \|}-\tilde{G} n(1-v_{\|})\right|, \ \ \Delta M=\frac{\Delta m^2 \cos 2 \theta}{4p^{\nu}_0}.
\end{aligned}
\end{equation}
The obtained expressions allow to consider \cite{Pustoshny:2018jxb} the effect of the resonance amplification of the neutrino spin-flavor oscillations in different background environments.
\color{Black}
\subsection*{Resonance amplification of neutrino spin-flavor oscillations $\nu_{e}^{L} \Leftarrow\left(j_{\perp}\right) \Rightarrow \nu_{\mu}^{R}$ by longitudinal matter current}

Here we examine \cite{Pustoshny:2018jxb, Shakhov:2019} the case when the
amplitude of oscillations $\sin ^{2} 2 \theta_{\mathrm{eff}}$ in (\ref{prob}) is not small
and we use the criterion based on the demand that
$\sin ^{2} 2 \theta_{\mathrm{eff}} \geq \frac{1}{2}$
which is provided by the condition $E_{\mathrm{eff}} \geq \Delta_{\mathrm{eff}}.$

Consider the case when the effect of the magnetic field is
negligible, thus we get
\begin{equation}
\left|\left(\frac{\eta}{\gamma}\right)_{e \mu} \tilde{G} n v_{\perp}\right| \geq|\Delta M-\tilde{G} n(1-v_{||})|.
\end{equation}
In the further evaluations we use the approximation
$\left(\frac{\eta}{\gamma}\right)_{e \mu}~\approx~\frac{\sin 2 \theta}{\gamma_{\nu}},$
where $\gamma_{\nu}= \gamma _{11}\sim \gamma _{22}$ is the neutrino effective gamma-factor.
In the case $v_{||}=0$ we get
\begin{equation}
\frac{\tilde{G} n v_{\perp}}{\gamma_{\nu}} \sin 2 \theta+\tilde{G} n \approx \tilde{G} n
\end{equation}
Finally, the criterion $\sin ^{2} 2 \theta_{\mathrm{eff}} \geq \frac{1}{2}$ is fulfilled when the following
condition is valid:
$\tilde{G} n \geq \Delta M.$

%\subsection*{Models of short gamma-ray bursts}
Consider a model of short gamma-ray bursts (see \cite{Grigoriev:2017wff, Perego:2014fma}). Neutrino escaping the central neutron star with inclination given by an angle $\alpha$ from the plane of the accretion disk propagates through the toroidal bulk of very dense matter that rotates with the angular velocity of about $\omega=10^3$ $\mathrm{s}^{-1}$ around the axis that is perpendicular to the
accretion disk. The diameter of the perpendicular cut of the toroidal bulk of matter is about $d \sim 20$ km and the distance
from the center of this cut to the center of the neutron star is also about $D \sim 20$ km. The transversal velocity of matter
can be estimated accordingly $v_{\perp}=\omega D=0.067$ that corresponds to $\gamma _n=1.002.$

The mass squared difference and mixing angle are taken from the solar neutrino measurements,
$\Delta m^2~=~7.37~\times~10^{-5}~\mathrm{eV}^{2},$ $\sin ^{2} \theta~=~0.297\ (\cos 2 \theta~=~0.406).$
Consider neutrino with energy $p^{\nu}_0=10^6$ eV and moving matter characterized by $\gamma _n=1.002$. Thus, we get $\Delta M=0.75 \times 10^{-11} \mathrm{eV}.$ Accounting for the estimation $\tilde{G}=\frac{G_{F}}{2 \sqrt{2}}=0.4 \times 10^{-23} \mathrm{eV}^{-2}$ from the criterion $\tilde{G} n \geq \Delta M$ we get the quite
reasonable condition on the density of neutrons,
%\begin{equation}
$n_{0} \geq \frac{\Delta M}{\tilde{G}}=10^{12} \mathrm{eV}^{3} \approx 10^{26}  \ \ \mathrm{cm}^{-3}$.
%\end{equation}
The corresponding oscillation length is approximately
\begin{equation}
L_{\mathrm{eff}}=\frac{\pi}{\left(\frac{\eta}{\gamma}\right)_{e \mu} \tilde{G} n v_{\perp}} \approx 5 \times 10^{11} \mathrm{km}.
\end{equation}
The oscillation length can be within the scale of short
gamma-ray bursts
%\begin{equation}
$L_{\mathrm{eff}} \approx 10 \ \  \mathrm{km}$
%\end{equation}
if the matter density equals $n_{0} \approx 5 \times 10^{36} \mathrm{cm}^{-3}.$
%
%\section*{Conclusions}

The performed studies \cite{Pustoshny:2018jxb, Shakhov:2019} of neutrino spin $\nu_{e}^{L} \Leftarrow\left(j_{\perp}\right) \Rightarrow \nu_{e}^{R}$ and spin-flavor $\nu_{e}^{L} \Leftarrow\left(j_{\perp}\right) \Rightarrow \nu_{\mu}^{R}$ oscillations engendered by the transversal matter currents in the presence of arbitrary magnetic fields allow one to consider possibilities of applications of these very interesting new effects in different astrophysical settings.

%----------------------------------------------------------------------------------------
%   REFERENCES
%----------------------------------------------------------------------------------------
\tiny

\end{document}